\documentclass[fleqn,usenatbib]{mnras}

\usepackage{newtxtext,newtxmath}

\usepackage[T1]{fontenc}

\DeclareRobustCommand{\VAN}[3]{#2}
\let\VANthebibliography\thebibliography
\def\thebibliography{\DeclareRobustCommand{\VAN}[3]{##3}\VANthebibliography}

\usepackage{graphicx}	
\usepackage{amsmath}	
\usepackage{ulem}

\title[Galaxy-velocity correlations in simulations]{Improving estimates of the growth rate using galaxy-velocity correlations: a simulation study}

\author[R. J. Turner et al.]{
Ryan J. Turner,$^{1}$\thanks{E-mail: rjturner@swin.edu.au}
Chris Blake,$^{1}$
Rossana Ruggeri$^{1}$
\\
$^{1}$Centre for Astrophysics \& Supercomputing, Swinburne University of Technology, P.O. Box 218, Hawthorn, VIC 3122, Australia\\
}

\date{Accepted XXX. Received YYY; in original form ZZZ}

\pubyear{2020}

\begin{document}
\label{firstpage}
\pagerange{\pageref{firstpage}--\pageref{lastpage}}
\maketitle

\begin{abstract}
We present an improved framework for estimating the growth rate of large-scale structure, using measurements of the galaxy-velocity cross-correlation in configuration space. We consider standard estimators of the velocity auto-correlation function, $\psi_1$ and $\psi_2$, the two-point galaxy correlation function, $\xi_{gg}$, and introduce a new estimator of the galaxy-velocity cross-correlation function, $\psi_3$.
By including pair counts measured from random catalogues of velocities and positions sampled from distributions characteristic of the true data,
we find that the variance in the galaxy-velocity cross-correlation function is significantly reduced. Applying a covariance analysis and $\chi^2$ minimisation procedure to these statistics, we determine estimates and errors for the normalised growth rate $f\sigma_8$ and the parameter $\beta = f/b$, where $b$ is the galaxy bias factor. We test this framework on mock hemisphere datasets for redshift $z < 0.1$ with realistic velocity noise constructed from the L-PICOLA simulation code, and find that we are able to recover the fiducial value of $f\sigma_8$ from the joint combination of $\psi_1$ + $\psi_2$ + $\psi_3$ + $\xi_{gg}$, with 15\% accuracy from individual mocks. We also recover the fiducial $f\sigma_8$ to within 1$\sigma$ regardless of the combination of correlation statistics used. When we consider all four statistics together we find that the statistical uncertainty in our measurement of the growth rate is reduced by $59\%$ compared to the same analysis only considering $\psi_2$, by $53\%$ compared to the same analysis only considering $\psi_1$, and by $52\%$ compared to the same analysis jointly considering $\psi_1$ and $\psi_2$.
\end{abstract}




\section{Introduction}
         
Peculiar velocities are the local velocities of galaxies relative to the motion expected from the general expansion of the universe.  These velocities, which may be directly measured for individual galaxies using standard-candle techniques, are imparted over cosmic time under the gravitational influence of large-scale structure, encoding information about gravitational perturbations and the growth rate of structure. This makes the peculiar velocity field a powerful probe of mass fluctuations over the largest scales \citep{Watkins2009, Feldman2010, Koda2014} and related peculiar velocity statistics an effective probe of gravitational physics, or (assuming a standard cosmological model) parameters such as the matter density $\Omega_m$ \citep{Ferreira1999}.

Peculiar velocity statistics such as the two-point peculiar velocity correlation functions ($\psi_1$ and $\psi_2$, \citealt{Gorski1989}), and their use in constraining cosmological parameters, have long been discussed \citep[e.g.,][]{Groth1989,Borgani2000,Wang2018,Dupuy2019}.  These statistics are measured from the observed radial component of peculiar velocities where, under standard assumptions, the radial peculiar velocity field carries the same information as the three-dimensional velocity correlation tensor \citep{Gorski1988}. Other studies have modelled and measured observed peculiar velocities using maximum-likelihood approaches \citep{Johnson2014,Huterer2017,Adams2020} or power-spectrum techniques \citep{Park2000,Park2006,Qin2019}.

Peculiar velocities in linear theory constrain a degenerate combination of the growth rate $f$ and $\sigma_8$, the amplitude of density fluctuations on scales of $8 \, h^{-1}$ Mpc. This combined parameter is referred to as the normalised growth rate of structure, $f\sigma_8$.  Different cosmological models predict different $f\sigma_8$ behaviours. In the standard Lambda Cold Dark Matter ($\Lambda$CDM) model of cosmology the growth rate is constant with scale, with predicted redshift behaviour $f = \Omega_m(z)^{0.55}$ \citep{Linder2005}.  Some theories of modified gravity, such as the $f(R)$ scenario \citep{Mirzatuny2019}, predict that the growth rate changes as a function of scale \citep{Baker2014}. The effects of modified gravity are only observed in fluctuations on the largest scales, where the peculiar velocities outperform other cosmological probes, making them an important tool in disambiguating cosmological models and testing the $\Lambda$CDM model of cosmology \citep{Koda2014,Howlett2017}.

Cross-correlations between peculiar velocities and the galaxy density field contain additional information about cosmological physics, which may be exploited in a joint analysis using models linking the density and velocity statistics \citep[e.g.,][]{Davis2011,Hudson2012,Carrick2015,Ma2015,Adams2017,Nusser2017,Boruah2019,Adams2020}.  The common sample variance between the velocity and density fields serves to significantly improve the accuracy with which key parameters may be determined \citep{Koda2014,Howlett2017}. 

Joint analyses of the galaxy velocity and density fields have often been implemented in a ``velocity-density comparison'' method \citep[e.g.,][]{Strauss1995,Carrick2015,Said2020} where the density field is used construct a model velocity field which is compared with peculiar velocity measurements at the locations of galaxies.  Such approaches most directly recover the parameter $\beta = f/b$, where $b$ is the linear galaxy bias factor which describes how the galaxy distribution traces the underlying mass distribution \citep{Kaiser1984}.  The resulting value of $\beta$ depends on the density-field tracer, and this approach is complementary to measurements using redshift-space distortions in the galaxy correlation function \citep[e.g.,][]{Hawkins2003}.  We choose to frame our analysis in terms of $\beta$ rather than $b\sigma_8$, due to the considerable literature in measurements of $\beta$.

Despite the potential of direct peculiar velocity measurements to test large-scale cosmological physics, the small sample sizes of current surveys have limited their potency as cosmological probes. However, future datasets, many with a specific focus on observational strategies to mitigate systematic errors in the measurements of velocity dispersions and of stellar populations along the fundamental plane, are rapidly increasing the competitiveness of peculiar velocities as a cosmological probe and already permitting measurements of the local growth rate or associated parameters with $10$-$20\%$ precision \citep[e.g.,][]{Davis2011,Hudson2012,Johnson2014,Carrick2015,Adams2017,Wang2018,Qin2019,Dupuy2019}. These measurements generally agree with early-time measurements of $\Omega_m$ and $\sigma_8$ from WMAP9 \citep{Bennett2013} and Planck \citep{PlanckCollab2018}, and redshift-space distortion measurements from local galaxy surveys \citep{Beutler2012}.

Current samples which have been used for peculiar velocity studies include the 6-degree Field Galaxy Survey \citep{Springob2014}, the {\it CosmicFlows} samples \citep{Tully2013,Tully2016} and local supernovae surveys \citep{Ganeshalingam2013,Krisciunas2017,Foley2018}.  Future datasets which may be utilised for peculiar velocity studies include the Taipan Galaxy Survey \citep{DeCunha2017}, the WALLABY HI Survey \citep{Koribalski2020}, the Dark Energy Spectroscopic Instrument \citep{DESICollab2016} and supernovae datasets such as the Zwicky Transient Facility \citep{Bellm2019}.

This work investigates the joint use of galaxy and peculiar velocity correlation statistics to constrain $f\sigma_8$ in simulated catalogues, in which we assume the density field and velocity field are measured from the same sample of objects.  We extend existing work by focusing on a joint analysis of auto- and cross-correlation statistics between galaxies and peculiar velocities in configuration space, breaking the degeneracy between $f\sigma_8$ and $b\sigma_8$ found in the cross-correlation function and improving constraints on cosmological parameters.  Furthermore, we present a new statistic $\psi_3$ that acts as an analogue to the mean pairwise velocity estimator $v_{12}$, and extend current estimators by introducing pair counts with random velocity catalogues, dramatically improving the variance in estimates of the galaxy-velocity cross-correlation function. Since our focus is on testing growth information present in the peculiar velocity field rather than redshift-space distortions in the density field, we do not include redshift-space distortions in this simulation study, but will return to this issue in future work.

The structure of the paper is as follows. Section \ref{sec:theory} describes the theory of the auto-correlation and cross-correlation functions used in this work. The cosmological simulation data we use and the method by which we apply our estimators to the data is described in Section \ref{sec:simulations}. Section \ref{sec:estimators} describes the derivations of the correlation function estimators. In Section \ref{sec:results} we describe how successful our methodology is at constraining estimates of the normalised growth rate. We conclude and discuss future plans for extensions to this work in Section \ref{sec:conclusions}.

\section{Theory}
\label{sec:theory}

\subsection{Velocity auto-correlation functions}
The general form of the two-point correlation tensor of the peculiar velocity field, which contains all the information about a Gaussian vector field, is
\begin{equation}
\Psi_{ij}(\vec{r}_A,\vec{r}_B) = \langle v_i(\vec{r}_A) \, v_j(\vec{r}_B) \rangle
\end{equation}
\citep{Gorski1988}, where $\vec{r}_A$ and $\vec{r}_B$ are two spatial positions, $v_i$ are the components of peculiar velocity, and $\langle ... \rangle$ represents the average measurement over different statistical realisations.  Assuming that the velocity field is irrotational, homogeneous and isotropic, and that velocity perturbations are linear, the velocity correlation tensor can be written as,
\begin{equation}
    \label{tensor}
    \Psi_{ij}(r) = [\Psi_\parallel(r) - \Psi_\perp (r)] \, \hat{r}_{Ai} \, \hat{r}_{Bj} + \Psi_\perp(r) \, \delta^K_{ij}
\end{equation}
where $r = | \vec{r}_B - \vec{r}_A |$ is the magnitude of the separation between positions $\vec{r}_A$ and $\vec{r}_B$, $\Psi_\parallel(r)$ and $\Psi_\perp (r)$ are functions describing the correlation between components of velocity parallel and perpendicular to the separation vector $\vec{r}$, and $\delta^K_{ij}$ is the Kronecker delta. The spectral form of $\Psi_\parallel(r)$ and $\Psi_\perp (r)$ was described by \cite{Gorski1988},
\begin{equation}
    \label{psipara}
    \Psi_\parallel(r) = \frac{H^2 a^2 (f\sigma_8)^2}{2\pi^2} \int \frac{P(k)}{\sigma_{8,{\rm fid}}^2} \left[ j_0(kr) - 2\frac{j_1(kr)}{kr} \right] dk
\end{equation}
\begin{equation}
    \label{psiperp}
    \Psi_\perp(r) = \frac{H^2 a^2 (f\sigma_8)^2}{2\pi^2} \int \frac{P(k)}{\sigma_{8,{\rm fid}}^2} \frac{j_1(kr)}{kr} dk
\end{equation}
where $H$ is the Hubble parameter, $P(k)$ is the linear matter power spectrum as a function of wavenumber $k$, which we assume in our study is measured at redshift $z = 0$ meaning the scale factor $a = 1$, and $j_i(x)$ is the i'th spherical Bessel function of the first kind,
\begin{equation}
    j_0(x) = \frac{\sin{x}}{x}
\end{equation}
\begin{equation}
    j_1(x) = \frac{\sin{x}}{x^2} - \frac{\cos{x}}{x}
\end{equation}
In this work we assume that the initial shape of the power spectrum on large scales is known, from Cosmic Microwave Background observations for example, and then assuming this shape consider measuring the amplitude of the velocity power (i.e., the growth rate of structure) in the late Universe.

Equations \ref{psipara} and \ref{psiperp} show the dependence of the parallel and perpendicular components of the velocity correlation tensor on the parameter $(f\sigma_8)^2$, after separating out a fiducial value of $\sigma_8$, highlighting the usefulness of these equations in constraining this combined parameter. We assume linear theory throughout this study (hence restrict our analysis to large scales), noting that extensions have been studied by \citet{Okumura2014}.

\begin{figure}
    \centering
    \includegraphics[width=\linewidth]{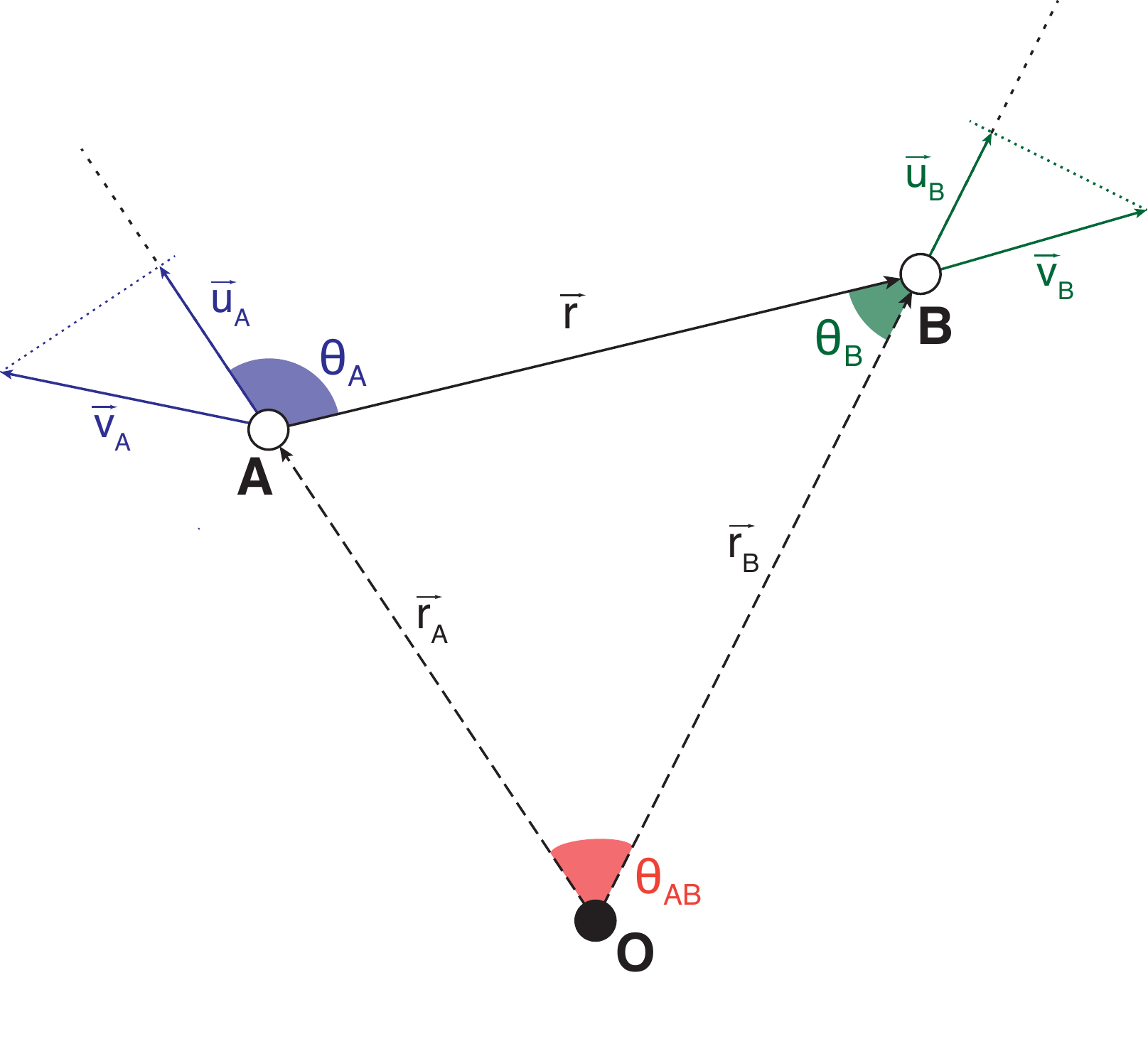}
    \caption{An example pair of galaxies A and B, as seen by an observer O, illustrating the geometry of the scenario.}
    \label{fig:geometry}
\end{figure}

As we can only measure the radial component of a galaxy's velocity in practice, the correlation tensor cannot be measured directly.  From equation \ref{tensor}, the correlation for line-of-sight velocities of two galaxies $u_A$ and $u_B$ with separation $\vec{r}$ can be written as,
\begin{equation}
    \label{uu_corr}
    \langle u_A(\vec{x})~u_B(\vec{x}+\vec{r})\rangle = \Psi_\perp \cos{\theta_{AB}} + [\Psi_\parallel - \Psi_\perp] \cos{\theta_A} \cos{\theta_B}
\end{equation}
where (with reference to Figure \ref{fig:geometry}), the angles between the galaxies are $\cos{\theta_{AB}} = \hat{\vec{r_A}} \cdot \hat{\vec{r_B}}$, $\cos{\theta_{A}} = \hat{\vec{r}} \cdot \hat{\vec{r_A}}$, and $\cos{\theta_{B}} = \hat{\vec{r}} \cdot \hat{\vec{r_B}}$.

\cite{Gorski1989} expressed the functions $\Psi_\parallel(r)$ and $\Psi_\perp (r)$ in terms of $\psi_1$ and $\psi_2$, velocity correlation statistics that depend on the radial peculiar velocity, which are described by,
\begin{equation}
    \label{psi1}
    \psi_1(r) \equiv \frac{\Sigma \,w_A w_B\,u_A u_B\, \cos{\theta_{AB}}}{\Sigma \,w_A w_B\, \cos^2{\theta_{AB}}}
\end{equation}
\begin{equation}
    \label{psi2}
    \psi_2(r) \equiv \frac{\Sigma \,w_A w_B \,u_A u_B\, \cos{\theta_{A}}\cos{\theta_{B}}}{\Sigma \,w_A w_B\, \cos{\theta_{AB}}\cos{\theta_{A}}\cos{\theta_{B}}}
\end{equation}
where the sums are taken over pairs of galaxies in a separation bin around $r$.  In the case of a pair of galaxies $A$ and $B$, at positions $\vec{r}_A$ and $\vec{r}_B$ with peculiar velocities $\vec{v}_A$ and $\vec{v}_B$, the radial component of peculiar velocity is $\vec{u}_{A,B}$ = ($\vec{v}_{A,B} \cdot \hat{\vec{r}}_{A,B}$)$ \cdot \hat{\vec{r}}_{A,B}$. These quantities are expressed visually in Figure \ref{fig:geometry}. $w_{A,B}$ are galaxy-specific weights related to the error in velocity measurement, and are discussed more in Section \ref{sec:weight}.
The numerators of equations \ref{psi1} and \ref{psi2} sum over the dot product of the radial peculiar velocities, and the product of the components of the radial peculiar velocities along the separation vector, respectively. The denominators in these equations normalise the sums such that the norm of the velocity field is preserved.

Using equation \ref{uu_corr} the models for $\psi_1$ and $\psi_2$ can be expressed as a function of both $\Psi_\parallel(r)$ and $\Psi_\perp(r)$,
\begin{equation}
\begin{aligned}
    \langle \psi_{1}(r) \rangle & = \frac{\Sigma \,w_A w_B\, \langle u_Au_B \rangle \cos{\theta_{AB}}}{\Sigma \,w_A w_B\, \cos^2{\theta_{AB}}} \\ 
                                & = \mathcal{A}(r)\Psi_\parallel(r) + [1 - \mathcal{A}(r)]\Psi_\perp(r)
\end{aligned}
\end{equation}
\begin{equation}
\begin{aligned}
    \langle \psi_{2}(r) \rangle & = \frac{\Sigma \,w_A w_B\, \langle u_A u_B \rangle \cos{\theta_{A}}\cos{\theta_{B}}}{\Sigma \,w_A w_B\, \cos{\theta_{AB}}\cos{\theta_{A}}\cos{\theta_{B}}} \\ 
                                & = \mathcal{B}(r)\Psi_\parallel(r) + [1 - \mathcal{B}(r)]\Psi_\perp(r)
\end{aligned}
\end{equation}
where $\mathcal{A}$ and $\mathcal{B}$ are functions describing the geometry of the survey, dictating the contributions of $\Psi_\parallel$ and $\Psi_\perp$ to $\psi_1$ and $\psi_2$, respectively,
\begin{equation}
    \mathcal{A}(r) = \frac{\Sigma \,w_A w_B\, \cos{\theta_A} \cos{\theta_{B}} \cos{\theta_{AB}}}{\Sigma \,w_A w_B\, \cos^2{\theta_{AB}}}
\end{equation}
\begin{equation}
    \mathcal{B}(r) = \frac{\Sigma \,w_A w_B\, \cos^2{\theta_A} \cos^2{\theta_{B}}}{\Sigma \,w_A w_B\, \cos{\theta_A} \cos{\theta_{B}} \cos{\theta_{AB}}}
\end{equation}
Analyses of the peculiar velocity correlation functions using these statistics can be found in several previous studies including \cite{Borgani2000}, \cite{Wang2018} and \cite{Dupuy2019}.

\subsection{Galaxy-velocity cross-correlation function}

We now extend our models to encompass the cross-correlation between the peculiar velocity and galaxy density fields (see also, \citealt{Nusser2017} and \citealt{Adams2017}).  The cross-correlation between the velocity and density fields is given by,
\begin{equation}
    \xi_{gv}(r) \, \hat{r} = \langle \delta(\vec{x}) \, \vec{v}(\vec{x} + \vec{r}) \rangle
\end{equation}
where,
\begin{equation}
    \xi_{gv}(r) = -\frac{Ha (f \sigma_8) (b \sigma_8)}{2\pi^2}\int dk \, k \, \frac{P(k)}{\sigma_{8,{\rm fid}}^2} \, j_1(kr)
\label{xigv}
\end{equation}
(see \citealt{Fisher1995, Adams2017}). 

The cross-correlation function between the line-of-sight velocity (at position $B$) and galaxy overdensity (at position $A$) is hence:
\begin{equation}
    \label{crosscorrfunc}
    \langle \delta_A(\vec{x}) \, u_B(\vec{x} + \vec{r}) \rangle = \xi_{gv}(r) \, \cos{\theta_B}
\end{equation}
where $\theta_B$ is defined in Figure \ref{fig:geometry}.  Thus, for a given galaxy-velocity pair separated by $r$, an estimator of $\xi_{gv}(r)$ is $u_B/\cos{\theta_B}$. By summing over many such pairs in this separation bin, taking a weighted mean across those pairs using inverse-variance weighting $w_\theta = 1/\sigma^2$ where $\sigma \propto 1/\cos{\theta_B}$ is the error in each individual estimate due to the varying angle with respect to the line of sight, we form an estimator for the galaxy-velocity cross-correlation function that is also dependent on radial peculiar velocities, which we call $\psi_3$,
\begin{equation}
\psi_3(r) \equiv \frac{\Sigma \,w_\theta w_B\, ~(u_B/\cos{\theta_B})}{\Sigma \,w_\theta w_B\,} = \frac{\Sigma \,w_B\,u_B \cos{\theta_B}}{\Sigma \,w_B\, \cos^2{\theta_B}}
\label{psi3}
\end{equation}
where $\langle \psi_3(r) \rangle = \xi_{gv}(r)$, $w_{\theta}$ is given above and $w_B$ is the same weight applied in Equations \ref{psi1} and \ref{psi2}, discussed further in Section \ref{sec:weight}.  This estimator follows the same structure as $\psi_1$ and $\psi_2$ from \cite{Gorski1989}.  The numerator of equation \ref{psi3} sums the component of the peculiar velocity of galaxy $B$ along the separation vector towards galaxy $A$ for all pairs of galaxies.  The $\cos^2{\theta_B}$ that appears in the denominator due to the $w_{\theta}$ weighting also preserves the norm of the velocity field, similarly to the denominators in equations \ref{psi1} and \ref{psi2}.  We note that $\psi_3$ is equivalent to the mean pairwise velocity estimator $v_{12}$ which can be measured from observational catalogues using an estimator introduced by \citet{Ferreira1999},
\begin{equation}
    v_{12} = \frac{2\Sigma(u_A + u_B)(\cos{\theta_A} + \cos{\theta_B})}{\Sigma(\cos{\theta_A} + \cos{\theta_B})^2}
\end{equation}

\subsection{Spatial two-point correlation function}

To complete the set of galaxy-velocity correlations we use in this study we also consider galaxy clustering, which is useful for constraining the galaxy bias, breaking the degeneracy between the bias and the growth rate in the galaxy-velocity cross-correlation in Equation \ref{xigv}.  The galaxy auto-correlation function is defined as,
\begin{equation}
    \xi_{gg}(r) = \langle \delta(\vec{x})~\delta(\vec{x} + \vec{r}) \rangle
\end{equation}
and measures the tendency for galaxies to cluster under the influence of gravity at a fixed separation $\vec{r}$.  The linear-theory model for $\xi_{gg}$ is,
\begin{equation}
    \xi_{gg}(r) = \frac{(b \sigma_8)^2}{2\pi^2}\int dk \, k^2 \, \frac{P(k)}{\sigma_{8,{\rm fid}}^2} \, j_0(kr) \, e^{-k^2a^2}
\end{equation}
where the term $e^{-k^2a^2}$ with $a = 1 \, h^{-1}$ Mpc does not affect the large-scale correlation function but ensures more efficient numerical convergence.

We measure the correlation function on large scales, and must therefore include the baryon acoustic oscillation (BAO) peak in our model.  This peak is not well-described by a linear power spectrum model, because it is smeared out by the motion of galaxies. This effect can be represented by modifying the linear power spectrum model \citep{Anderson2012} as,
\begin{equation}
    P(k) = P_{\rm nw}(k)\left[1 + \left(\frac{P_{\rm lin}(k)}{P_{\rm nw}(k)} - 1\right)e^{-\frac{1}{2}k^2\Sigma_{nl}^2}\right]
\end{equation}
where $P_{\rm lin}(k)$ is the linear power spectrum of our cosmological model and $P_{\rm nw}(k)$ is a no-wiggles matter power spectrum model created using formulae from \cite{Eisenstein1998} in which the BAO peak is removed, and $\Sigma_{nl}$ is a parameter describing the damping of the acoustic peak due to galaxy displacement. We set $\Sigma_{nl} = 10 \, h^{-1}$ Mpc for this analysis, typical for low-redshift and ensuring a good description of our data.

\section{Simulations}
\label{sec:simulations}

We investigate our correlation function estimators using simulations, in order to compare the different estimators defined in Section \ref{sec:estimators} and demonstrate that they recover unbiased cosmological parameters.  For this analysis we use dark matter halo catalogues generated for the Taipan Galaxy Survey project \citep{DeCunha2017} using the L-PICOLA \citep{Howlett2015} N-body simulation code in a box of scale-length $1800 \, h^{-1}$ Mpc.  L-PICOLA is a fast, distributed-memory, planar-parallel code based on Co-moving Lagrangian Acceleration (COLA, \citealt{Tassev2013}) that generates dark matter fields.  By taking fewer time steps in the simulation process, COLA is able to perform faster than full N-body simulations, at the cost of poorer resolution on small scales. Velocity statistics are dominated by large-scale modes, and so worse accuracy on small scales is not a concern for this work \citep{Koda2016}. See \citet{Howlett2015} for information on how L-PICOLA calculates velocity and position values.

The mocks were constructed from initial conditions corresponding to a fiducial cosmological model with parameters $\Omega_m = 0.3121$, $\Omega_b = 0.0491$, $n_s = 0.9653$, $h = 0.6751$ and $\sigma_8 = 0.8150$, and built from snapshots generated at redshift $z=0$.  At this redshift, the fiducial value of the parameter $f\sigma_8 = 0.4296$.  We select 30 of these mocks, containing dark matter halos with masses in the range $12.4 < \log_{10}(h^{-1} \, M_\odot) < 12.6$. Our results are not sensitive to the range of halo mass used for this study, given that velocity bias is $\sim 1$ on large scales \citep{Desjacques2010}.

We subdivide these mocks into 36 hemispherical regions with radius $300 \, h^{-1}$ Mpc, for a total of 1080 independent datasets that share no common halos. This geometry matches a typical wide-area observational survey across a hemisphere to $z = 0.1$, such as the 6-degree Field Galaxy Survey or the Taipan Galaxy Survey. Each of these mocks initially contain approximately 24 000 halos, and we select a subsample of $N_D =$ 10 000 halos before proceeding with our analysis, matching the approximate size of current PV datasets.  The random datasets needed for this analysis, as outlined in Section 3, are populated with $N_R =$ 50 000 halos.  The observers in each of these datasets are placed at the centre of the hemisphere's face along the $y$-$z$ plane, and radial peculiar velocities are calculated for each dataset with respect to the central observer. 

We introduce a random error to our velocities in order to mimic the scaling of the observational error with distance that is seen in survey data. We sample errors for each galaxy from a normal distribution with standard deviation $\sigma = H_0 d \cdot e = 100h \cdot d \cdot e$, where $\sigma$ has units of km s$^{-1}$, $d$ is the distance to the galaxy from the observer in $h^{-1}$ Mpc and $e$ is some value describing the fractional error in the measurement of the distance. We take $e$ to be 0.15, representative of the accuracy of distance measurements from the Tully-Fisher relation.  We do not consider any other observational effects -- such as redshift space distortions -- at this time, in order to isolate the information arising from the velocity field, and its cross-correlation with the density field.

We perform a growth-rate analysis of individual mocks, and we also analyse sets of ``stacked'' mocks where we reduce sample variance by averaging correlation function measurements over groups of 20 realisations.  Individual mocks are representative of results from current PV surveys, and mock averages allow us to verify that our linear-theory model for the galaxy and velocity correlation functions holds at an accuracy significantly better than required, and to test our conclusions with a higher degree of precision.  We average over 20 mocks as it produces a significantly more accurate representation than existing data samples, whilst still being susceptible to accurate modelling on large scales by our linear-theory representation

\begin{figure*}
    \centering
    \includegraphics[width = \textwidth]{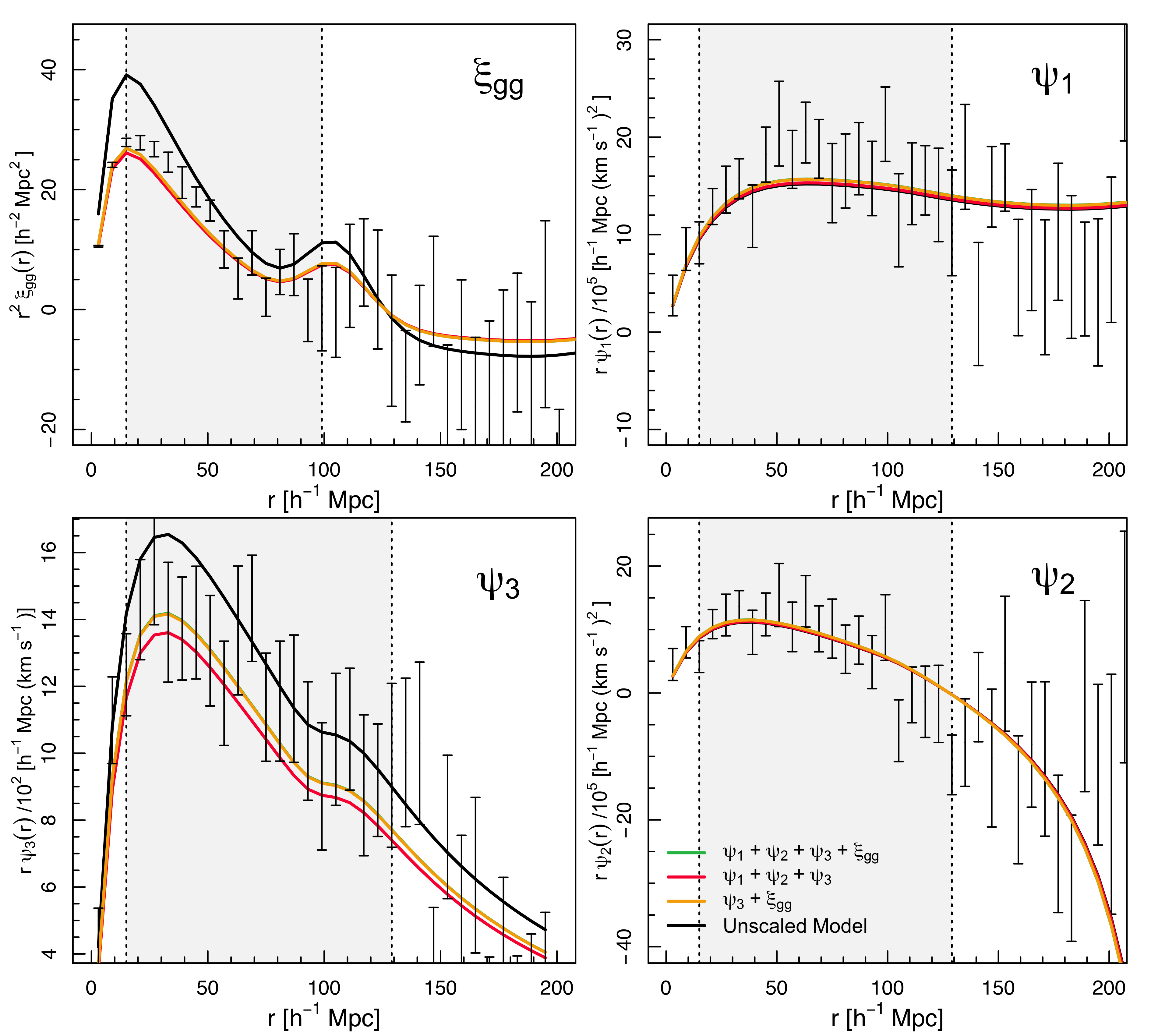}
    \caption{An example of parameter fitting for each of the four statistics considered in this work. The unscaled model for each statistic is shown in black, and the model rescaled using the best-fitting $f\sigma_8$ and $b\sigma_8$ (calculated from $\beta$) from the combinations ($\psi_1 + \psi_2 + \psi_3 + \xi_{gg}$), ($\psi_1 + \psi_2 + \psi_3$) and ($\psi_3 + \xi_{gg}$) are shown in different colours given in the legend in the top left panel. Errors represent the standard deviation in the measured values across one sample of 20 mocks. Shaded regions in each panel depict the parts of the data used to fit for $f\sigma_8$ and $b\sigma_8$. Top left: $\xi_{gg}$, multiplied by the separation squared $r^2$. Top right: $\psi_1$, multiplied by $r/10^5$. Bottom left: $\psi_3$, multiplied by the separation $r/10^2$. Bottom right: $\psi_2$, multiplied by $r/10^5$. Note that the green line representing the joint four correlation statistic case is often obfuscated by the yellow line representing the ($\psi_3 + \xi_{gg}$) case.}
    \label{fig:estimators}
\end{figure*}

For one average of 20 randomly chosen mocks, Figure \ref{fig:estimators} shows each of the models unscaled and recalculated using the best-fitting parameters derived from fitting three combinations of correlation statistics to the amplitude of the galaxy and velocity auto- and cross-correlation functions: ($\psi_3 + \xi_{gg}$), ($\psi_1 + \psi_2 + \psi_3$) and ($\psi_1 + \psi_2 + \psi_3 + \xi_{gg}$). The shaded regions in Figure \ref{fig:estimators} indicate the range used to fit for $f\sigma_8$ and $b\sigma_8$. We fit the ranges $15 < r < 129 \, h^{-1}$ Mpc for $\psi_1, \psi_2$ and $\psi_3$, and $15 < r < 99 \, h^{-1}$ Mpc for $\xi_{gg}$, in bins of width $6 \, h^{-1}$ Mpc. We exclude the very smallest scales, restricting ourselves to larger scales for which linear-theory modelling is expected to apply, and choose the upper limits for our fitting ranges by testing different values and using those which optimise our final results for signal-to-noise and the stability of the resulting covariance matrix.

In Figure \ref{fig:estimators} the unscaled models for each statistic are shown in black and each differently coloured line represents a model calculated from a set of best-fitting cosmological parameters derived from different combinations of statistics.  Whilst the inclusion or exclusion of individual statistics has some influence on the best-fitting models, the overall amplitudes of $f\sigma_8$ and $b\sigma_8$ are statistically consistent, as we will discuss in Section \ref{sec:results}.


Measurements in different separation bins and between different statistics will be correlated, owing to the common sample variance and the fact that individual galaxies join pairs in multiple separation bins. Accounting for this correlation requires the covariance between our statistics to be considered across all separations when fitting for model parameters. 

A joint fit to multiple statistics is necessary to break the degeneracies between the parameters $f\sigma_8$ and $b\sigma_8$ (or $\beta$).  To construct the covariance matrix between the different statistics and separation bins, we first concatenate the sequence of statistics used in the analysis to form a total data vector $A(i)$.  We use the 1080 realisations to compute the covariance between our statistics.  The covariance between two bins $i$ and $j$ is measured by,
\begin{equation}
    C_{ij} = \frac{1}{N - 1} \sum_{k = 1}^{N} \left[ A_k(i) - \overline{A(i)} \right] \left[ A_k(j) - \overline{A(j)} \right]
\end{equation}
where $N$ is the number of realisations and $\overline{A(i)}$ and $\overline{A(j)}$ are the mock means of the statistics for bins $i$ and $j$, respectively.  We scale the resulting covariance to match the number of realisations forming our mock means.

\begin{figure}
    \centering
    \includegraphics[width = \linewidth]{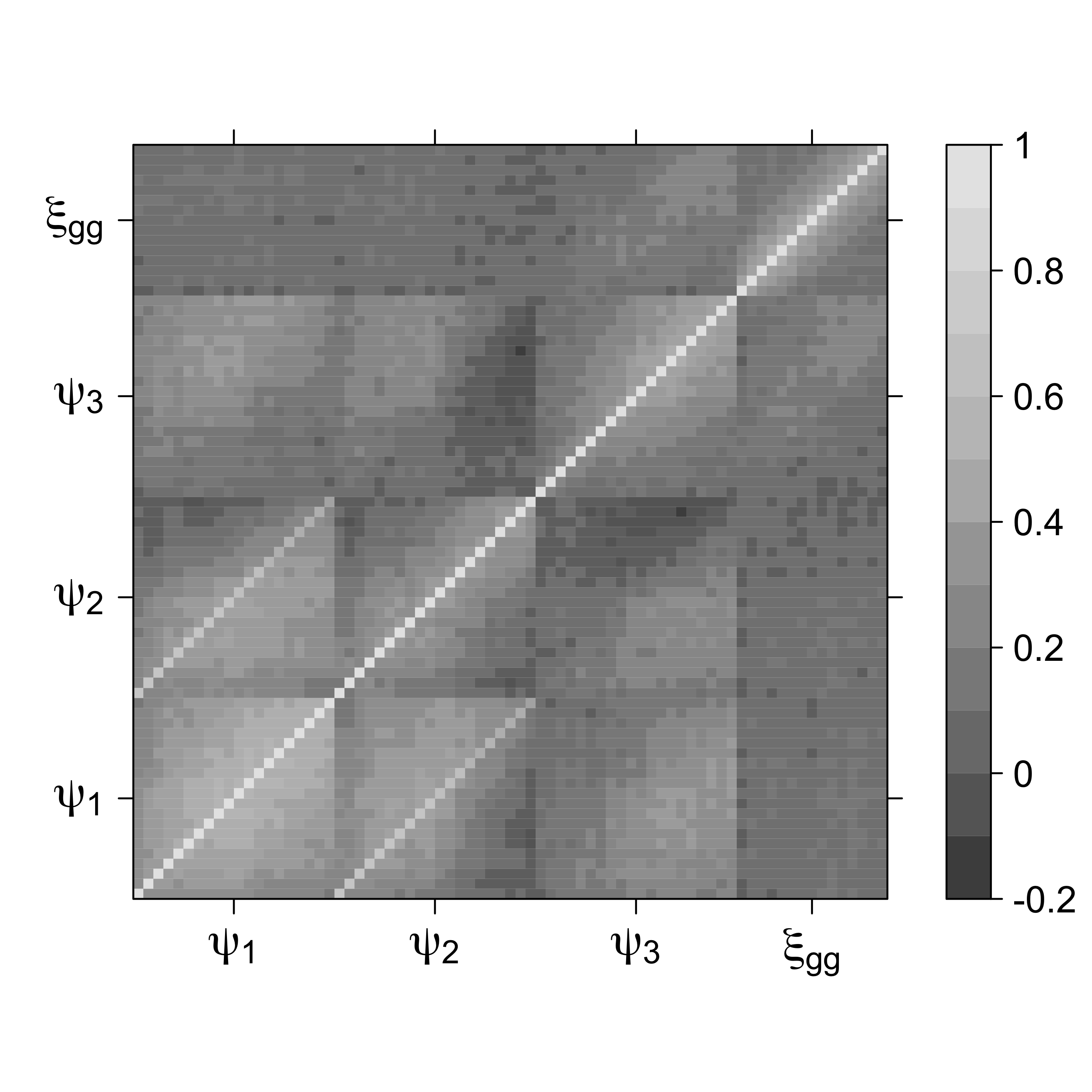}
    \caption{Reduced covariance matrix, dimensions 75 $\times$ 75, for the ($\psi_1 + \psi_2 + \psi_3 + \xi_{gg}$) combination of statistics. Each cell corresponds to one separation bin in the fitting range of each statistic, given in Section \ref{sec:results}, increasing from left to right and bottom to top. The diagonal elements of the covariance matrix are unity by definition, and are shown in white.}
    \label{fig:covariance}
\end{figure}

The correlation matrix for the vector $A = [ \psi_1(r), \psi_2(r), \psi_3(r), \xi_{gg}(r) ]$ is shown in Figure \ref{fig:covariance}.  Along the diagonal, when we measure the correlation between the measurement of statistic $A$ in bin $i$ and itself, we see perfect correlation. The velocity correlation functions $\psi_1$ and $\psi_2$ are reasonably correlated between statistics and scales, shown by the faint diagonals in the ($\psi_1$, $\psi_2$) and ($\psi_2$, $\psi_1$) sections of the matrix, but the correlation in the off diagonal terms is mitigated by the velocity errors we select. The minimum correlation coefficient in this region of our reduced covariance matrix in Figure \ref{fig:covariance} is $\approx -0.12$, when correlating $\psi_1$ measurements at the smallest scales with $\psi_2$ measurements at the largest scales.  The off-diagonal correlations involving $\psi_3$ are lower, but there is some correlation between $\psi_3$ and $(\psi_1, \psi_2)$. $\xi_{gg}$ has the lowest amplitude of cross-correlation with the other statistics.

\section{Correlation function estimators}
\label{sec:estimators}

We measure the four statistics $\psi_1, \psi_2, \psi_3, \xi_{gg}$ discussed above using configuration-space estimators. We use a combination of a position and peculiar velocity mock dataset of $N_D$ galaxies and a corresponding randomly-generated dataset of $N_R$ points. The randomly generated data are distributed over the same volume, with positions sampled from the same distribution as the mock data and uncorrelated peculiar velocities sampled from a normal distribution with variance including both sample variance ($290$ km/s, matching the distribution of the mock data) and measurement noise ($\sigma$ introduced in Section \ref{sec:simulations}) -- which we term `random velocities'. These datasets can be used to construct different pair counts at separations $r$, which can be used to estimate our ensemble of statistics. We attribute weights to these randomly-distributed galaxies using the method described in Section \ref{sec:weight}.

Starting with the galaxy correlation function, a basic, unbiased estimator of $\xi_{gg}$ was introduced by \cite{Peebles1974},
\begin{equation}
    \hat{\xi}_{gg}(r) = \frac{N_R^2}{N_D^2}\frac{D_g D_g(r)}{R_g R_g(r)} - 1
    \label{shortform_xigg}
\end{equation}
combining the $D_g D_g(r)$ and $R_g R_g(r)$ galaxy-galaxy pair counts from our dataset and random catalogues, respectively, in a separation bin around $r$.  The error in this estimator can be reduced by introducing the pair count $D_g R_g$, representing the cross-pairs between the data and random catalogues, to form the new estimator \citep{Landy1993},
\begin{equation}
    \hat{\xi}_{gg}(r) = \frac{N_R^2}{N_D^2}\frac{D_g D_g(r)}{R_g R_g(r)} - 2\frac{N_R}{N_D}\frac{D_g R_g(r)}{R_g R_g(r)} + 1
    \label{longform_xigg}
\end{equation}
This extension reduces the variance in the estimator by lowering the statistical error in the measurement associated with the distribution of data points with respect to the sample boundaries.

\begin{figure*}
    \centering
    \includegraphics[width=\textwidth]{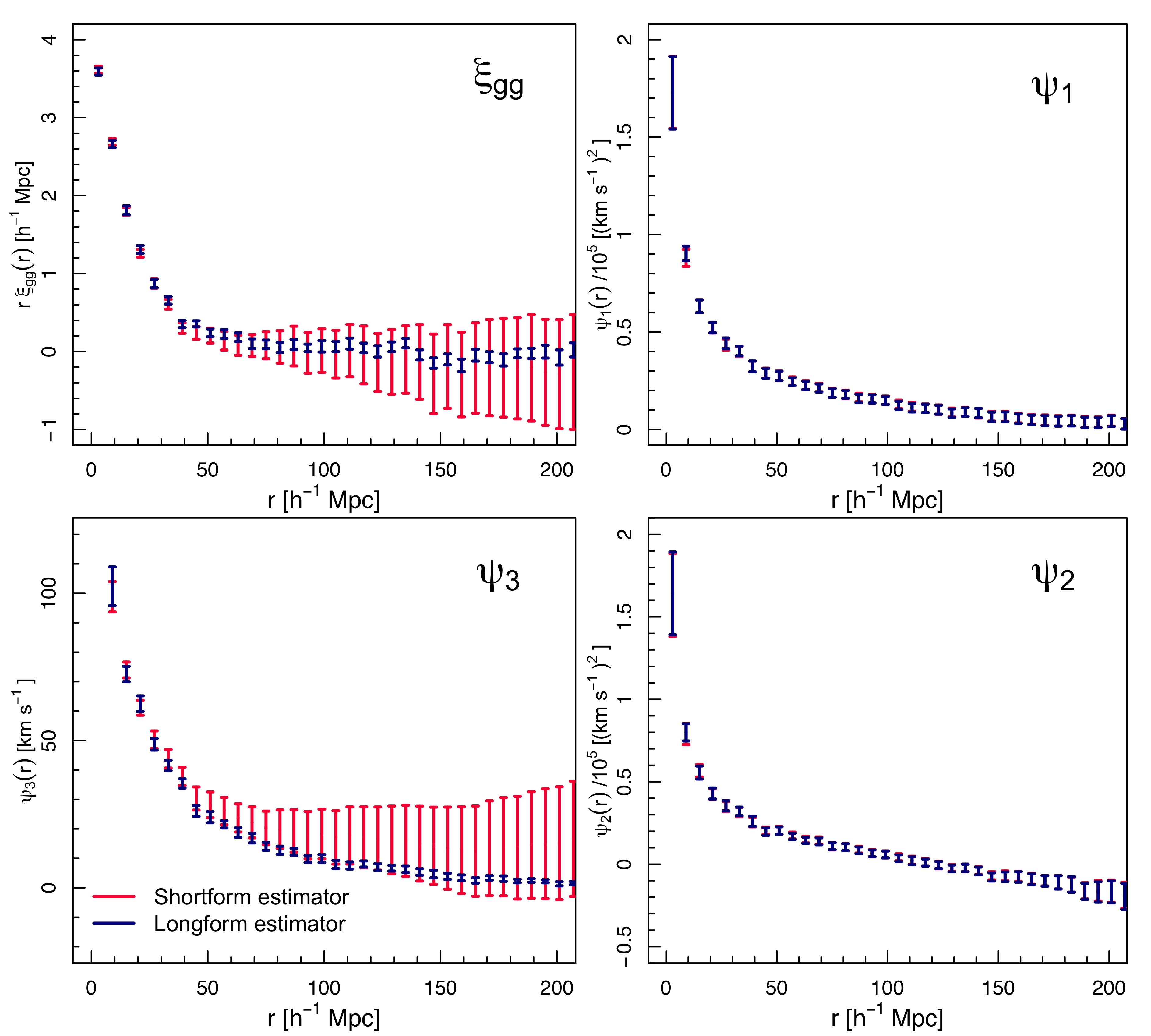}
    \caption{Comparison between the short form of the correlation function estimators (red), evaluated using equations \ref{shortform_xigg}, \ref{shortform_xigv} and \ref{shortform_xivv}, and the long form of the estimators including additional random pair counts (blue), evaluated using equations \ref{longform_xigg}, \ref{longform_xigv} and \ref{longform_xivv}. Errors represent the standard deviation in the measured values across one sample of 20 mocks, and are centred on the mean of the measured values. The variance of the short-form and long-form estimators is very similar for $\psi_1$ and $\psi_2$, the velocity auto-correlation functions, but much larger reductions are seen in $\xi_{gg}$ and $\psi_3$, which have a dependence on scale. Top left: $\xi_{gg}$, multiplied by the separation $r$ for clarity of display. Top right: $\psi_1$, multiplied by $1/10^5$. Bottom left: $\psi_3$. Bottom right: $\psi_2$, multiplied by $1/10^5$.}
    \label{fig:long_v_short}
\end{figure*}

This argument can be extended to estimators involving the line-of-sight peculiar velocity components as well as position components. A simple short-form estimator of the galaxy-velocity cross-correlation function is
\begin{equation}
    \label{shortform_xigv}
    \hat{\psi}_3(r) = \frac{N_R^2}{N_D^2}\frac{DD_{\psi_{3,n}}(r)}{RR_{\psi_{3,d}}(r)}
\end{equation}
where $DD_{\psi_{3,n}} = \sum_{A,B} \,w_b\, u_B \cos{\theta_B}$ is the numerator of the $\psi_3$ estimator shown in Equation \ref{psi3}, evaluated for cross-pairs between data galaxies $A$ and data velocities $B$, and $RR_{\psi_{3,d}} = \sum_{A,B} \,w_B\, \cos^2{\theta_B}$ is the denominator of the same $\psi_3$ estimator, evaluated for random galaxies $A$ and random velocities $B$.  This estimator can be extended by including random position and random velocity components, written in the same notation as above:
\begin{equation}
    \label{longform_xigv}
    \begin{aligned}
    \hat{\psi}_3(r) = &\frac{N_R^2}{N_D^2}\frac{DD_{\psi_{3,n}}(r)}{RR_{\psi_{3,d}}(r)} - \frac{N_R}{N_D}\frac{DR_{\psi_{3,n}}(r)}{RR_{\psi_{3,d}}(r)} \\
                        &- \frac{N_R}{N_D}\frac{RD_{\psi_{3,n}}(r)}{RR_{\psi_{3,d}}(r)} + \frac{RR_{\psi_{3,n}}(r)}{RR_{\psi_{3,d}}(r)}
    \end{aligned}
\end{equation}
This extended estimator has a similar effect of reducing variance in the measured cross-correlation function (see Figure \ref{fig:long_v_short} and the discussion below).

Likewise, the short-form and long-form estimators for $\psi_{1,2}$ have the form,
\begin{equation}
    \label{shortform_xivv}
    \hat{\psi}_{1,2}(r) = \frac{N_R^2}{N_D^2}\frac{DD_{\psi_{(1,2),n}}(r)}{RR_{\psi_{(1,2),d}}(r)}
\end{equation}
and,
\begin{equation}
    \label{longform_xivv}
    \hat{\psi}_{1,2}(r) = \frac{N_R^2}{N_D^2}\frac{DD_{\psi_{(1,2),n}}(r)}{RR_{\psi_{(1,2),d}}(r)} - \frac{N_R}{N_D}\frac{DR_{\psi_{(1,2),n}}(r)}{RR_{\psi_{(1,2),d}}(r)} + \frac{RR_{\psi_{(1,2),n}}(r)}{RR_{\psi_{(1,2),d}}(r)}
\end{equation}
where the notation $DD_{\psi_{(1,2),(n,d)}}$ refers to the pair count associated with the numerator ($n$) or denominator ($d$) of the estimator for the $\psi_1$ statistic (Equation \ref{psi1}) or the $\psi_2$ statistic (Equation \ref{psi2}), evaluated for data-data ($DD$), data-random ($DR$) or random-random ($RR$) pairs.  In Section \ref{sec:results} we consider the relative performance of these different estimators.

The normalisation of the $\xi_{gg}$ correlation function requires knowledge of the true number density of galaxies, but this quantity can only be estimated from the data itself.  This results in an additive bias to the measured correlation function known as the integral constraint correction \citep{Peebles1974b, Peebles1980, Landy1993, Scranton2002}.

\begin{equation}
I.C. = \frac{\sum_i \xi_{gg}(r_i) \, R_gR_g(r_i)}{\sum_i R_gR_g(r_i)}
\end{equation}
where the sum is taken over all separation bins to the maximum which may be embedded in the survey, and we take the $\xi_{gg}$ term in the numerator as our fiducial model. We find that I.C. $= 2.56 \times 10^{-4}$ for our hemispherical geometry, which we add to our measurement of $\xi_{gg}$. As the expectation value of the average velocity is zero, there is no analogous integral constraint for the galaxy-velocity or velocity-velocity correlations.

\subsection{Weighting}
\label{sec:weight}
As well as introducing extensions to the estimators, we also need to include a weight for each object in our estimators, due to the distance-dependent errors associated with each galaxy. The minimum-variance weight has the form
\begin{equation}
w = \frac{1}{P_v \cdot n_g + \sigma^2}
\end{equation}
similar to a weight chosen by \citealt{Qin2019} based on work done by \citealt{Feldman1994}, where $P_v$ is the value of the velocity power spectrum at a desired scale in units of $(h^{-1} {\rm Mpc})^3 ({\rm km}\, {\rm s}^{-1})^2$, $n_g$ is the number density of galaxies in the sample in units of $(h^{-1} {\rm Mpc})^{-3}$ and $\sigma$ is the galaxy-specific standard deviation, in units of km s$^{-1}$, used to derive the velocity error in Section \ref{sec:simulations}. We set $n_g = 10^{-3}$ (h$^{-1}$ Mpc)$^{-3}$ to match our mock catalogues, and find that a value of $P_v = 10^9 \, h^{-3}$ Mpc$^3$ km$^2$ s$^{-2}$, which is characteristic of the model velocity power at scale $k \approx 0.05 \, h$ Mpc$^{-1}$, produces the most accurate measurement of the resulting growth rate (after trialling different choices for this quantity). This weight is applied to each galaxy contributing to the PV estimator. The weights we apply to each galaxy scale with their error and up-weight galaxies with more accurately measured peculiar velocities, which improves our measurements on small scales where random noise would otherwise dominate. Conversely large-scale measurements, where cosmic variance dominates, are made slightly noisier by the inclusion of the weight as each galaxy no longer contributes to the measurement equally.

\section{Results}
\label{sec:results}

\begin{figure}
    \centering
    \includegraphics[width = \linewidth]{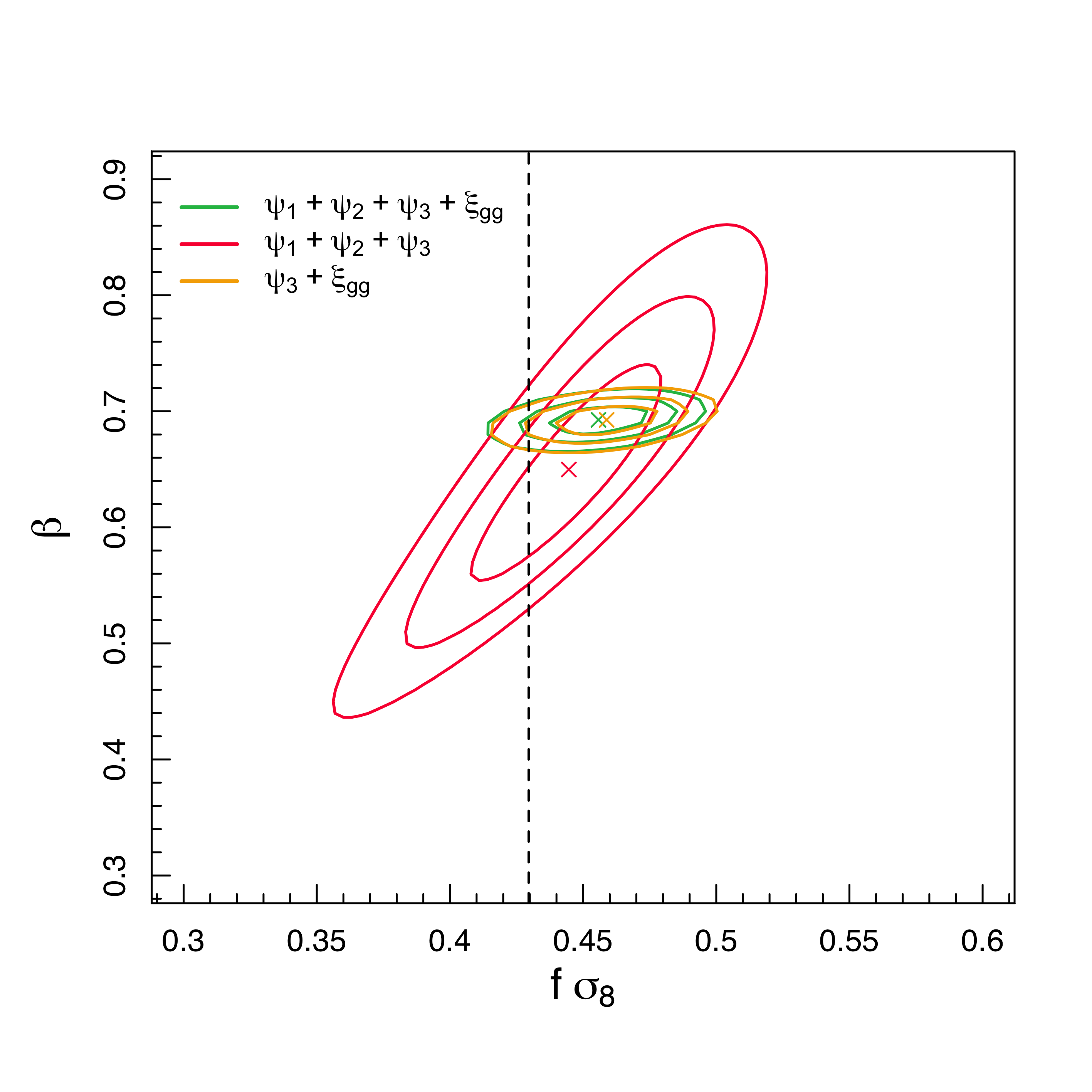}
    \caption{Contours in $f\sigma_8 - \beta$ parameter space for combinations ($\psi_1 + \psi_2 + \psi_3 + \xi_{gg}$), ($\psi_1 + \psi_2 + \psi_3$) and ($\psi_3 + \xi_{gg}$), derived from the same 20-mock sample used in Figure \ref{fig:1Dfits}. Contours represent $1\sigma$, $2\sigma$ and $3\sigma$ confidence intervals.}
    \label{fig:2Dfits}
\end{figure}

\begin{figure}
    \centering
    \includegraphics[width = \linewidth]{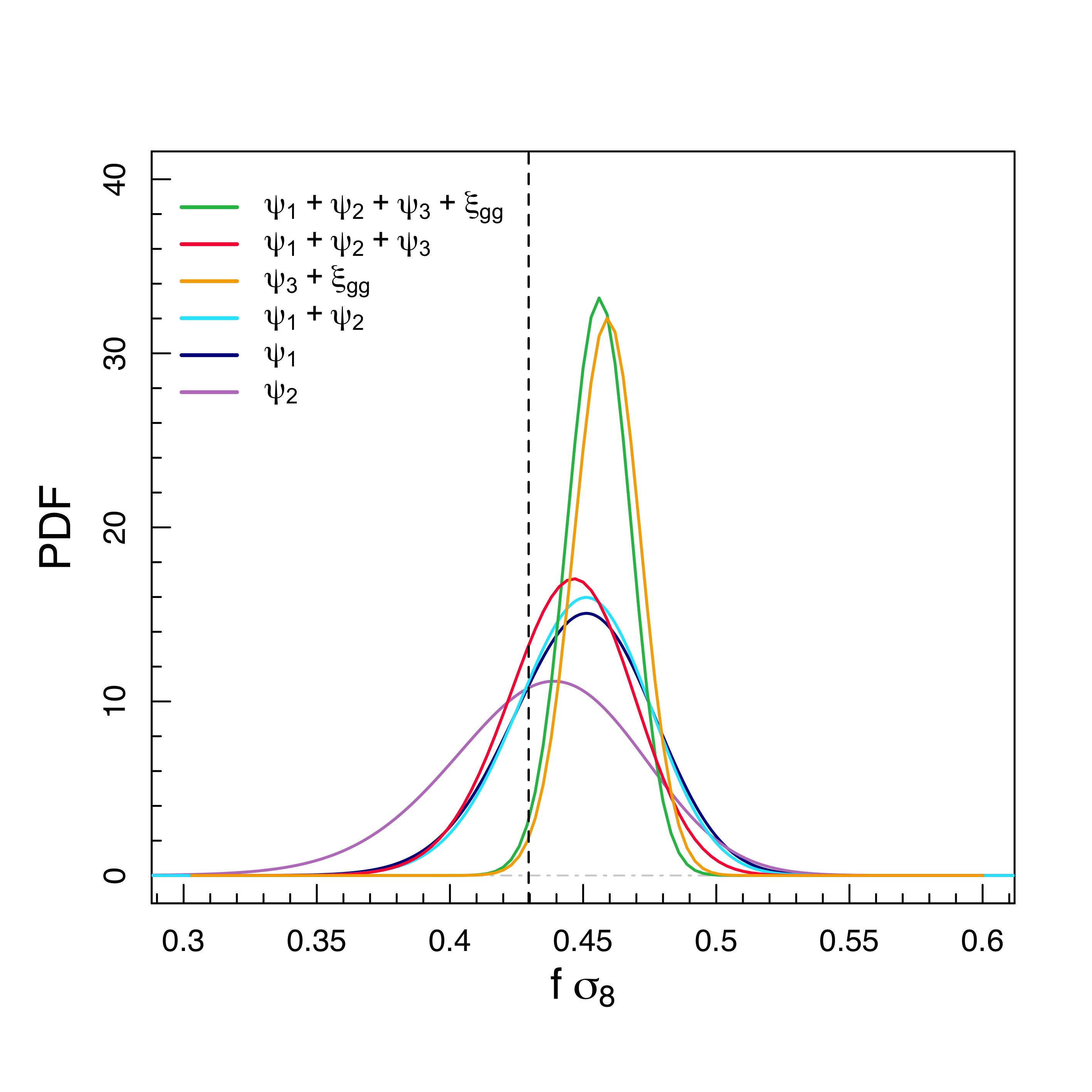}
    \caption{One-dimensional posterior distributions for $f\sigma_8$ from a sample of 20 mocks using various different treatments of the statistics considered in this work. ($\psi_1 + \psi_2 + \psi_3 + \xi_{gg}$), ($\psi_1 + \psi_2 + \psi_3$) and ($\psi_3 + \xi_{gg}$) are dependent on both $f\sigma_8$ and $\beta$ through $b\sigma_8$, and so must be marginalised to recover the 1D posterior. The addition of $\psi_3$ improves the PDF, and the addition of $\xi_{gg}$ improves the PDF further.}
    \label{fig:1Dfits}
\end{figure}

We test our estimators and models by determining the best-fitting values of the normalised growth rate $f\sigma_8$ and parameter $\beta$ for our mock datasets, from which we can also produce a best-fitting value of the galaxy bias factor $b\sigma_8$ to input into our models for $\psi_3$ and $\xi_{gg}$. We generate fiducial models using the input cosmological parameters of the simulation, and rescale these models using the trial parameters based on the dependencies described in Section \ref{sec:theory}.  We use a $\chi^2$ minimisation procedure to obtain the best-fitting $f\sigma_8$ and $\beta$ values by fitting the models for $\psi_1$, $\psi_2$, $\psi_3$ and $\xi_{gg}$ against our measurements of those statistics.  The best-fitting value of the model is obtained by minimising
\begin{equation}
    \label{eq:chisq}
    \begin{aligned}
    \chi^2(f\sigma_8, \beta) = \sum_{i,j = 1}^N ~& (A_d(i) - A_m(i; f\sigma_8, \beta)) ~C_{ij}^{-1}\\
                                           & (A_d(j) - A_m(j; f\sigma_8, \beta))
    \end{aligned}
\end{equation}
where $A_{d,m}$ represents the concatenation of statistics for the data and model, respectively. The likelihood function of each of these models is proportional to $\exp{(-\chi^2/2)}$.  We recover $f\sigma_8$ and $\beta$ from joint parameter fits (Figure \ref{fig:2Dfits}) by marginalising the two-dimensional posterior probability distribution to find separate constraints for each parameter (Figure \ref{fig:1Dfits}). We apply these fits both to individual mocks, and averages over groups of 20 mocks.  This enables us to both study the precision of the fits for representative current PV samples, as well as test that our conclusions hold at a higher level of precision.  To understand the dispersion of our results across different samples of 20 mocks, we repeat the best-fitting procedure 1000 times, in each case randomly selecting a new sample of 20 mocks from the overall ensemble of 1080 L-PICOLA hemispheres.

\begin{figure*}
    \centering
    \includegraphics[width = \linewidth]{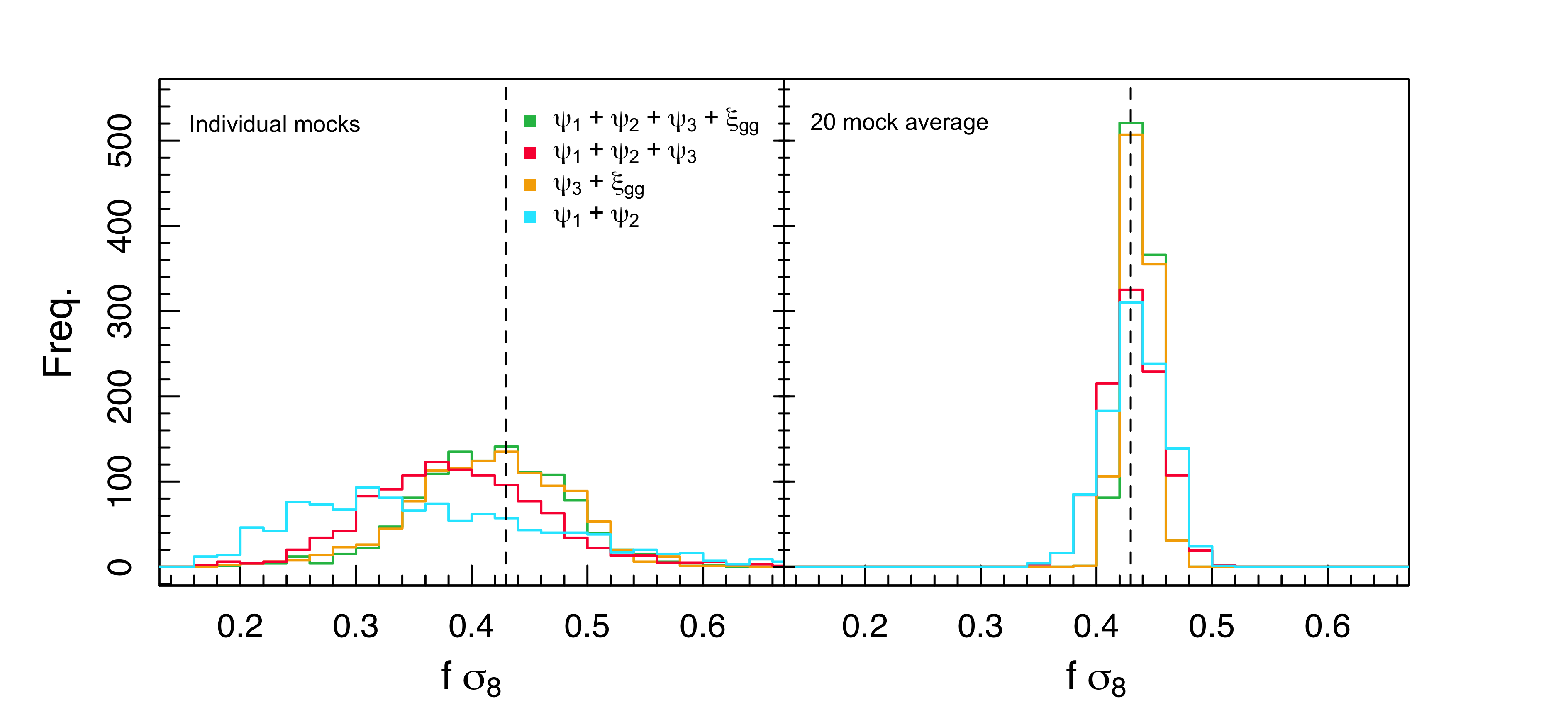}
    \caption{Distribution of recovered $f\sigma_8$ values, calculated from marginalised joint posteriors such as those shown in Figure \ref{fig:1Dfits}, for ($\psi_1 + \psi_2 + \psi_3 + \xi_{gg}$), ($\psi_1 + \psi_2 + \psi_3$), ($\psi_3 + \xi_{gg}$) and ($\psi_1 + \psi_2$) from all 1080 individual mocks (\textit{left}) and from all 1000 samples of 20 random mocks (\textit{right}). The combination of statistics ($\psi_3 + \xi_{gg}$) is able to constrain $f\sigma_8$ well despite its weaker dependence on the growth rate, but ($\psi_1 + \psi_2 + \psi_3 + \xi_{gg}$) is the strongest combination. In the individual mock results, the inclusion of $\psi_3$ noticeably improves the results, as can be seen in the difference between the red and blue histograms.}
    \label{fig:fsig_dist}
\end{figure*}

The distribution of values of $f\sigma_8$ recovered from the marginalised posteriors such as those in Figure \ref{fig:1Dfits} is shown in Figure \ref{fig:fsig_dist} for individual mocks (left-hand panel) and groups of 20 averaged mocks (right-hand panel). The fits to individual mocks span a larger range than the 20-mock averaged results, as is to be expected, and the effect of including $\psi_3$ is clearly seen when comparing the results in blue for ($\psi_1 + \psi_2$) to the results in red for ($\psi_1 + \psi_2$ + $\psi_3$).  The results from the right side of Figure \ref{fig:fsig_dist} are in good agreement with the posteriors shown in Figure \ref{fig:1Dfits}, which we would expect.

In Table \ref{tab:fsigTable} we present the mean $f\sigma_8$ and $\beta$ values, and their corresponding $1\sigma$ confidence intervals, recovered from the posteriors of 1000 20-mock samples and 1080 individual L-PICOLA mocks -- both shown in Figure \ref{fig:fsig_dist} -- for various combinations of correlation statistics.  We also report the degrees of freedom involved in each combination and the corresponding average reduced $\chi^2$. The values for $\psi_1$, $\psi_2$ and $\psi_1 + \psi_2$ are found from a one-dimensional posterior, as these statistics only depend on $f\sigma_8$. The other combinations include some dependence on $\beta$ through $b\sigma_8$, and so must be found by marginalising over the two-dimensional posterior. The $\chi^2$ values typically indicate that the models are a good fit to the data, and the recovered growth rates are generally consistent with the fiducial value of $f\sigma_{8,{\rm fid}} = 0.4296$.

The one-dimensional posteriors for an example 20-mock sample are shown in Figure \ref{fig:1Dfits}. It can be seen that the $f\sigma_8$ posteriors are narrowed by the introduction of $\psi_3$, and narrowed further when considering $\xi_{gg}$ in conjunction. By showing the mean values recovered from the 20-mock samples, as well as for the individual mocks, it can be seen that the improvement in our results gained by the inclusion of $\psi_3$ and $\xi_{gg}$ is present in both the individual mock and average-mock analyses.

The same data presented in Figure \ref{fig:1Dfits} are also shown in Figure \ref{fig:2Dfits}, which shows the joint fit in $f\sigma_8 - \beta$ parameter space for the three combinations of statistics that are dependent on both.  In this particular case the three combinations produce best-fitting parameters which all recover the fiducial value of $f\sigma_8$, depicted by a vertical dashed line, with varying degrees of accuracy. ($\psi_3 + \xi_{gg}$) and ($\psi_1 + \psi_2 + \psi_3$) present visually different contours, and agree with each other on a 1$\sigma$ level. The combination of all four statistics, ($\psi_1 + \psi_2 + \psi_3 + \xi_{gg}$), agrees with the other combinations on a 1$\sigma$ level and presents comparatively tighter constraints than either combination.

We now consider the accuracy with which different combinations of statistics are able to recover the growth rate.  We consider here the individual-mock fits, although the 20-mock average results are similar, and the two cases may be compared in Table \ref{tab:fsigTable}. The error in the measurements from the individual mocks is roughly a factor of $\sqrt{20}$ larger than in the corresponding 20-mock average case, which is to be expected. In the case of our individual-mock fits, $\psi_1$ and $\psi_2$ used separately are able to constrain the value of the normalised growth rate to $f\sigma_8 = 0.3664 \pm 0.1358$ and $f\sigma_8 = 0.3666 \pm 0.1548$, respectively. $\psi_2$ alone places weaker constraints on $f\sigma_8$ than if we were to use $\psi_1$, producing an average error that is approximately $14\%$ larger. When used together, however, ($\psi_1 + \psi_2$) are able to predict a value of $f\sigma_8 = 0.3679 \pm 0.1312$ -- producing constraints which are $3\%$ tighter when compared to $\psi_1$ and $15\%$ tighter when compared to $\psi_2$.  For our chosen configuration, ($\psi_3 + \xi_{gg}$) is also a robust probe of $f\sigma_8$ despite its weaker dependence on the parameter, predicting $f\sigma_8 = 0.4144 \pm 0.0663$. The addition of these statistics to the ($\psi_1 + \psi_2$) fit further improves the accuracy of our measurement to $f\sigma_8 = 0.4151 \pm 0.0632$, producing the smallest error of any considered combination and recovering $f\sigma_8$ with 15\% accuracy.

Presenting our results in this manner also allows us to comment on the efficacy of $\psi_2$. This was originally discussed in \citet{Gorski1989} where $\psi_2$ was introduced, in which it was stated that the statistic was unstable when applied to datasets of the time and was subsequently dropped from further analysis. This is supported by \citet{Dupuy2019}, who state that $\psi_2$ is not robust enough to estimate $f\sigma_8$ on \textit{cosmicflows} type catalogues, but challenged by \citet{Wang2018} who state that $\psi_2$ is well-behaved on such catalogues. Using our methodology over the fitting range outlined in Section \ref{sec:simulations} we can recover the fiducial cosmology to within 1$\sigma$ using $\psi_2$, indicating its robustness on catalogues similar to those we use in our analysis.

\begin{table*}
    \centering
    \begin{tabular}{l|c|c|c|c|c|c|c}
    \hline
                                   &         &  & \multicolumn{1}{|l|}{Individual Mocks} & &  & 20-mock Average  & \\
    Statistic(s)                   &   $\nu$ & $\langle f\sigma_8 \rangle \pm \langle 1\sigma \rangle$ & $\langle \beta \rangle \pm \langle 1\sigma \rangle$ & $\langle \chi^2_\nu \rangle$ & $\langle f\sigma_8 \rangle \pm \langle 1\sigma \rangle$ & $\langle \beta \rangle \pm \langle 1\sigma \rangle$ & $\langle \chi^2_\nu \rangle$\\
    \hline
    $\psi_1$                                &   19 & $0.3664 \pm 0.1358$ & -                     & 1.0005 & $0.4316 \pm 0.0279$ & -                     & 1.0409\\
    $\psi_2$                                &   19 & $0.3666 \pm 0.1548$ & -                     & 1.0006 & $0.4217 \pm 0.0381$ & -                     & 0.9872\\
    $\psi_1 + \psi_2$                       &   39 & $0.3679 \pm 0.1312$ & -                     & 1.0002 & $0.4336 \pm 0.0262$ & -                     & 1.0041\\
    $\psi_3 + \xi_{gg}$                     &   33 & $0.4144 \pm 0.0663$ & $0.6835 \pm 0.0379$   & 1.0018 & $0.4362 \pm 0.0130$ & $0.6873 \pm 0.0084$   & 1.0888\\
    $\psi_1 + \psi_2 + \psi_3$              &   58 & $0.3872 \pm 0.0954$ & $0.6252 \pm 0.2091$   & 1.0031 & $0.4313 \pm 0.0243$ & $0.6677 \pm 0.0674$   & 1.0156\\
    $\psi_1 + \psi_2 + \psi_3 + \xi_{gg}$   &   73 & $0.4151 \pm 0.0632$ & $0.6836 \pm 0.0364$   & 1.0012 & $0.4366 \pm 0.0124$ & $0.6876 \pm 0.0080$   & 1.0416\\
    \hline
    \end{tabular}
    \caption{Mean $f\sigma_8$ and $\beta$ values and errors from 1000 $\chi^2$ analyses of different combinations of statistics measured from 20 randomly chosen mocks from our 1080 total L-PICOLA mocks, as well as the mean values and errors as measured from each of those 1080 mocks individually. We also report the degrees of freedom $\nu$ for each combination of statistics, and corresponding average reduced $\chi^2$, $\langle \chi^2_\nu \rangle$.}
    \label{tab:fsigTable}
\end{table*}

\section{Conclusions}
\label{sec:conclusions}
In this work we have developed a joint framework for studying galaxy and peculiar velocity correlation functions in configuration space, and tested this framework using accurate mock datasets.  We have particularly considered the galaxy-velocity cross-correlation function, introducing a new correlation function estimator $\psi_3$ that is analogous to the mean pairwise velocity estimator $v_{12}$, written in a similar formalism to the velocity auto-correlation estimators $\psi_1$ and $\psi_2$ introduced by \cite{Gorski1989}.  We have also investigated the form of these estimators, demonstrating that the variance of the cross-correlation estimator may be significantly reduced by including pair counts against random catalogues in a similar manner to galaxy auto-correlation functions \citep{Landy1993}. We also increase the accuracy of our measurements at small scales by introducing a weight to the estimators which scales with the error in the measurement of peculiar velocity for each galaxy.

Using $\psi_3$ and our improved cross-correlation estimators, alongside established peculiar velocity statistics $\psi_1$ and $\psi_2$, we measure the value of combined parameters $f\sigma_8$ and $\beta$ from large-scale cosmological simulation halo catalogues generated in the L-PICOLA framework.  We compute the covariance between our statistics using 1080 L-PICOLA hemispherical mocks, across a range of separations, and measure the average values and errors of $f\sigma_8$ and $\beta$ over each of the 1080 L-PICOLA mocks. Using this method we are able to successfully recover the intrinsic L-PICOLA value of $f\sigma_8 = 0.4296$ to within 1$\sigma$ for all combinations of statistics, and when using the joint combination of ($\psi_1$ + $\psi_2$ + $\psi_3$ + $\xi_{gg}$) we measure $f\sigma_8 = 0.4151 \pm 0.0632$. Applying this framework to all four correlation statistics available to us, we are able to recover the fiducial L-PICOLA growth rate with 15\% accuracy. We repeat this process and measure $f\sigma_8$ and $\beta$ from an ensemble of 1000 different samples of 20 mocks randomly chosen from our 1080 mocks, and find again that we are able to recover the fiducial f$\sigma_8$ to within 1$\sigma$, while also reducing the error in our measurement by approximately a factor of $\sqrt{20}$. In this case we measure $f\sigma_8 = 0.4366 \pm 0.0124$ from the joint consideration of all four statistics, recovering the fiducial L-PICOLA growth rate with 2.8\% accuracy.

The best-fitting parameters that we derive from various combinations of these statistics are able to accurately model the data on large scales.  While the velocity auto-correlation estimators $\psi_1$ and $\psi_2$ can accurately recover the normalised growth rate $f\sigma_8$, both by themselves and when used in conjunction with one another, the addition of the cross-correlation estimator $\psi_3$ and two-point spatial correlation function $\xi_{gg}$ adds further information to these fits, which reduces the errors in our measurements. By extending the considered statistics from just ($\psi_1 + \psi_2$) to ($\psi_1 + \psi_2 + \psi_3 + \xi_{gg}$) we can obtain an average error reduction in the individual mock $f\sigma_8$ measurement of approximately $52\%$ without impacting the average reduced $\chi^2$. A larger improvement is seen in the average error of $\beta$, as the two statistics impacted by our new, extended estimators $\psi_3$ and $\xi_{gg}$ contain dependencies on the combined parameter $b\sigma_8$, which we can measure from our recovered values of $f\sigma_8$ and $\beta$. Considering all four correlation statistics rather than just ($\psi_3 + \xi_{gg}$) or ($\psi_1 + \psi_2 + \psi_3$) reduces the average error in $\beta$ by approximately $4\%$ and $83\%$, respectively. Similar reductions are also seen in the 20-mock averaged measurements of $f\sigma_8$ and $\beta$.

In future work we intend to extend this framework to include redshift-space distortions quantified using the correlation function multipoles, and apply our method to datasets from existing and upcoming redshift and peculiar velocity surveys, such as Cosmicflows-3, 6dFGS and the Taipan Galaxy Survey.  This will result in accurate measurements of cosmological parameters from galaxy-velocity correlations, hence tests of the gravitational physics of the late-time Universe.

\section*{Acknowledgements}

We would like to thank Michael Strauss and Roohi Dalal for their careful reading of the paper and for their insightful comments, which have made this a stronger manuscript. RJT would like to acknowledge the financial support received through a Swinburne University Postgraduate Research Award throughout the creation of this work. We thank Cullan Howlett and Pascal Elahi for generating the PICOLA simulations for the Taipan Galaxy Survey project.  We have used R \citep{Rteam2019} for our data analysis, and acknowledge that the plots in this paper were generated with the use of the {\sc  \textbf{magicaxis}} package \citep{magicaxis2019}. We also acknowledge additional funding through Australian Research Council Discovery Project DP160102705.

\section*{Data Availability}
The data underlying this article will be shared on reasonable request to the corresponding author.



\bibliographystyle{mnras}
\bibliography{galvel_corr} 





\bsp	
\label{lastpage}
\end{document}